\documentstyle[prl,aps,epsf]{revtex} % run with TeX 3.1415

\begin{document}

%\draft

%\preprint{IPT-99-}

\title
{ Solution of classical stochastic one dimensional many-body systems }

\author
{P.-A. Bares and M. Mobilia}

\address
{
Institut de Physique Th\'{e}orique,
\'{E}cole Polytechnique F\'{e}d\'{e}rale de Lausanne,
CH-1015 Lausanne 
}

\date{July , 1999 }

\maketitle

\begin{abstract}

We propose a simple method that allows, in one dimension, to solve exactly
a wide class of classical stochastic many-body systems far from equilibrium.
For the sake of illustration and without loss of generality,
we focus on a
model that describes the asymmetric diffusion of hard core particles in the
presence of an external source and instantaneous annihilation.  
Starting from a Master equation formulation of the problem we show 
that the density and multi-point correlation functions obey a closed set of
integro-differential equations which in turn can be solved numerically and/or
analytically.

\end{abstract}
\narrowtext
Recently, nonequilibrium statistical mechanics in low dimensional systems
has attracted much interest \cite{Privman97}. 
Whereas classical statistical mechanics has a well
established body of concepts and tools to deal with systems at or near
thermodynamic equilibrium, the situation is less satisfactory for far from
equilibrium systems.
The theoretical understanding of classical stochastic many-body systems where
both purely diffusive motion (due for example to static disorder, phonons, etc.)
and reactions (for example fusion, annihilation, coagulation, etc., where the
number of particles is not conserved)
take place simultaneously is relevant to a wide class of phenomena in physics,
biology, economics, a.s.o.\\
In view of this, we have chosen to investigate an important class of models
describing diffusion-limited reactions (for recent reviews see 
\cite{DLSchuetz99}). While it has long been recognized that the mean field 
approach is not 
applicable in
one and two dimensions, it is only 
recently that much progress has been achieved
in the description of the anomalous kinetics of low dimensional systems. 
Some open problems in this field are reminiscent of those 
studied intensively in the field
of strongly correlated low-dimensional systems. Hence, one might hope to use
with advantage the powerful tools devised in the latter.\\
The last decade has witnessed
new developments in the Master equation approach. The Markov generator of the
stochastic time evolution for the probability distribution can be interpreted 
as the Hamiltonian of quantum spin chains (with quantum group symmetries)
or alternatively of fermion or boson many-body systems with 
interaction terms. The dynamics is then
coded in an imaginary time Schr\"odinger equation or alternatively formulated as an
Euclidean field theory. The most formal 
approach expresses the Hamiltonian as a sum of generators of an Hecke
Algebra and, through Baxterization, associates an integrable 
vertex model \cite{Jones,Droz}.
The free fermion models considered by Grynberg et {\it al.} 
\cite{Grynbergetal}, Sch\"utz and others (see \cite{DLSchuetz99} and references
therein)
can be viewed as special cases of the above with a grading of the
Yang-Baxter algebra.
As is known in the field of integrable models, it is in general
extremely difficult, if not impossible, to extract the correlations from the Bethe
``wave functions'', even in their algebraic form (see for example 
\cite{Korepinetal}).
This is unfortunate because conceptually as well as experimentally (see
\cite{Privman97} and references therein), the
correlations coded in these ``wave functions'' are the most relevant quantities.
We should mention at this point however that there is one successful
application of ideas developed in the field of strongly correlated 
systems, namely  
the extension of 
the numerical Density Matrix Renormalization Group to nonequilibrium 
statistical mechanics
in one dimension\cite{DMrenorm99}.\\
On the analytical front, a powerful method is 
that developed by Cardy
and collaborators \cite{Cardyetal}. 
Assuming that the system is in a low-density regime, these authors rewrite the
Markov generator of the stochastic time evolution as a bosonic Hamiltonian 
which in turn can be treated by field theory and
renormalization group techniques in arbitrary dimension. For certain models, it
is even possible to evaluate exactly the beta functions for the coupling 
constants \cite{Cardyetal}. This is a first indication that
the underlying mathematical structure is substantially simpler than that of 
the quantum many-body problem. To our knowledge, this observation as well as 
its implications
do not seem to have drawn much
attention in the literature.\\
Despite the progress achieved in the last decade, there are 
still many open or poorly understood problems, even in one dimension: 
the multi-species and the 
epidemic models, the propagation of catalytic shock waves, etc. 
In response to these challenges, we were lead to consider a fermionic
field theory formulation of the Master equation for dense systems 
\cite{BaresMobilia99}. While inquiring
about the possible application of the renormalization group to the fermion field
theory, we realized that it
is possible to express the density and correlation functions as solutions of
a closed set of integro-differential equations. These in turn can be solved
exactly numerically and/or analytically. This is an unexpected and remarkable
result, especially in view of the large amount of work 
that has been performed in the field using sophisticated methods. Our approach 
in contrast is simple and elementary. In this letter we would like to report on
this rather unanticipated observation.\\ 
For illustrative purposes, we focus here on an extended Lushnikov model
\cite{Lushnikov}
that describes the (asymmetric) diffusion of particles in the presence of  
bilinear source and
annihilation. Certainly, such a model
does not constitute a protoptype for classical stochastic many-body systems
as does the Ising model for phase transitions: the
variety of phenomena to be described in nonequilibrium statistical mechanics 
is much too rich to be coded in a 
single model. Though we are perfectly aware of the numerous 
potential applications 
that our method provides, we will refrain from mentioning 
them here due to a lack of
space and refer the reader to elsewhere (e.g.
\cite{Privman97,DLSchuetz99}).\\
Consider a lattice of $L$ ($L/2$ even) sites with periodic boundary conditions
on which classical particles with hard core can diffuse with rates $h$ to the
left and $h'$ to the right, respectively. On encounter, pairs of particles 
annihilate with
rate $\epsilon'$. A source injects particle pairs with rate $\epsilon$. 
Rewriting the
Master equation as a Schr\"odinger equation in imaginary time, we have
$d|P(t)\rangle/dt=-H|P(t)\rangle$, with ($\sigma_{m+L}^+=\sigma_{m}^+$)\\
\begin{eqnarray}
\label{1}
H=-\sum_{m=1}^{L}\Bigg\{&&h\sigma_{m+1}^{+}\sigma_{m}^{-}
+h'\sigma_{m}^{+}\sigma_{m+1}^{-}+\epsilon'\sigma_{m}^{-}
\sigma_{m+1}^{-}+ \epsilon(\sigma_{m}^{+}\sigma_{m+1}^{+}-1)
\nonumber\\&+&(2\epsilon-h-h')\sigma_{m}^{+}\sigma_{m}^{-}
-(\epsilon'+\epsilon-h-h') \sigma_{m}^{+}\sigma_{m}^{-} 
\sigma_{m+1}^{+}\sigma_{m+1}^{-} \Bigg\}
\end{eqnarray}
\\
where $\sigma_{m}^{\pm}$ denote the standard Pauli matrices. 
The Dirac ket $|P(t)\rangle=
\sum_\eta P_{\eta}(t)|\eta\rangle$, where $\eta$ denotes a random configuration
in Fock space, encodes the probabilities of the various configurations
(see \cite{Grynbergetal}). The above dynamics is supplemented with
the initial conditions $|P(0)\rangle=|\rho\rangle$ (see below).  
It is convenient to
introduce the left vacuum $\langle{\widetilde\chi}|=\sum_{\eta}\langle\eta|$.
Conservation of probability implies then $\langle{\widetilde\chi}|H=0$, while
expectation values are linear in the kets $\langle A(t)\rangle =
\langle{\widetilde\chi}|A|P(t)\rangle$.
The models that have been solved previously correspond to 
$\epsilon'+\epsilon=h+h'$ 
(see e.g.\cite{DLSchuetz99,Grynbergetal,BaresMobilia99,Lushnikov}),
i.e., to quadratic hamiltonians. Below, we show that the Hamiltonian $H$ 
can be solved exactly for $\gamma=\epsilon'+\epsilon-(h+h')\not=0$ 
by elementary means, i.e., that the
density and two-point (multi-point) correlation function(s) 
can be computed exactly even though
the hamiltonian is non-quadratic.\\
Performing a Jordan-Wigner transformation, 
and subsequently a Fourier tranformation, 
leads to
\begin{eqnarray}
\label{2}
H&=&\sum_{q>0} \left[\omega(q) a_q^{\dagger}a_q +\ 
{\bar\omega}(q) a_{-q}^{\dagger}a_{-q} +2\sin{q}
\left(\epsilon'a_q a_{-q} + \epsilon a_{-q}^{\dagger}
a_{q}^{\dagger}\right)\right]+\epsilon L\nonumber\\  
&-&\gamma\sum_q (1-\cos q)a_q^{\dagger}a_q + 
\frac{\gamma}{L} \sum_{q_1,q_2,q_3} \cos(q_1-q_2) 
a_{q_1}^{\dagger}a_{q_2}a_{q_3}^{\dagger}a_{q_1+q_3-q_2}\; ,
\end{eqnarray}
where $\omega(q)=a-b\cos q+i(h'-h)\sin q$, with $a=\epsilon'-\epsilon,\; 
b=\epsilon'+\epsilon$. 
Following reference \cite{DLSchuetz99}, we consider a
translationally invariant and uncorrelated initial state 
with an even number $N$ of particles of density $\rho=N/L$, i.e.,
\begin{eqnarray}
\label{3}
|\rho\rangle=\frac{2(1-\rho)^L}{1+(1-2\rho)^L}\prod_{q>0} 
\Big(1-\mu^2 \cot(\frac{q}{2}) a_{-q}^{\dagger}
a_{q}^{\dagger}\Big)|0\rangle\; ,
\end{eqnarray}
where $\mu=\rho/(1-\rho)$.

\begin{figure}[b]
 \setlength{\unitlength}{1mm}
  \begin{picture}(80,80)
  \epsfxsize=7cm
 \epsfysize=7cm
 \put(35,0){\epsfbox{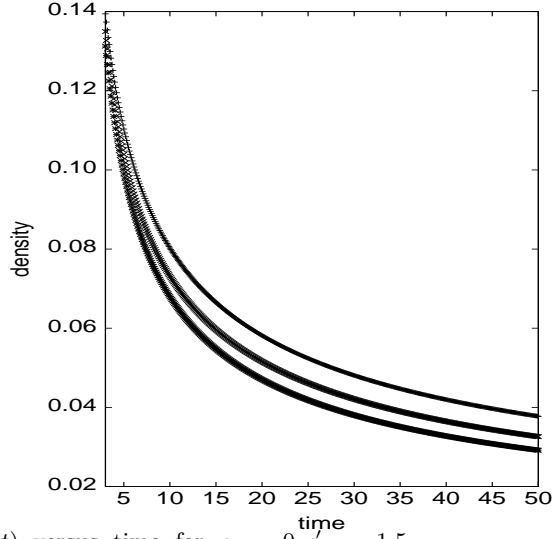}}
  \end{picture}
 \caption{Density $\rho(t)$ versus time for $\epsilon=0, 
 \epsilon'=1.5$ and 
 $\gamma=0.5$; (upper curve)\, $\gamma=0 $ 
 (middle curve: free-fermions);\,$\gamma=-0.5$ (lower curve).
 \label{fig1}}
\end{figure}

Using the properties of the left vacuum $(\langle \widetilde \chi| a_q^{\dagger}=\cot(\frac{q}{2})\langle \widetilde \chi| a_{-q}$, see \cite{DLSchuetz99}), and with the notation $\langle a_{-q}a_{q'}\rangle (t)\equiv \langle \widetilde \chi|a_{-q} a_{q'} e^{-Ht}|\rho\rangle $ , the density and the
two-point density correlation function can be written as 
\begin{eqnarray}
\label{5}
\rho(t)=\langle a_m^{\dagger}a_m\rangle(t)=\frac{1}{L}\sum_{q,q'}
e^{i(q-q')m}\cot(\frac{q}{2})\langle \widetilde\chi|a_{-q} a_{q'} 
e^{-Ht}|\rho\rangle
\end{eqnarray}
\begin{eqnarray}
\label{6}
\langle n_m n_n \rangle (t)=
\frac{1}{L^2}\sum_{k,k',q,q'}e^{im(q-q')+in(k-k')}
&\Bigg[&\langle a_{-q}a_{q'}\rangle(t)\delta_{k',k}
-\langle a_{q'}a_{k'}\rangle(t)\delta_{q,-k}
\nonumber\\&+&\cot(\frac{k}{2})\langle 
a_{-k}a_{-q}a_{q'}a_{k'}\rangle(t)\Bigg]\cot(\frac{q}{2})
\end{eqnarray}
Futhermore, it can be shown that \cite{BMStochastic99} 
\begin{eqnarray}
\label{7}
\langle \widetilde\chi|a_{-q} a_{q'} e^{-Ht}|\rho\rangle &=& 
g(q,t)\langle a_{-q}a_{q'}\rangle (0)=g(q,t)\frac{\mu^2 
\cot(\frac{q}{2})}{1+\mu^2 \cot(\frac{q}{2})^2}\delta_{q,q'},
\end{eqnarray}
where $g(q, t)$ is a function (even in $q$) that is a solution of a non-linear
integro-differential equation (see below),
with $g(q,t=0)=1$, $g(q,t<0)=0$ and $g(q,t)=g(q+2\pi, t)$.
This remarkable property can be generalized to higher order correlation 
functions. Due to the conservation of probability, an analog of the 
Wick theorem applies and all multipoint correlation functions can be 
computed.
For simplicity's sake we consider in the following a lattice initially 
full, i.e., $\rho=1$ (for the general case see \cite{BMStochastic99}) 
and obtain 
the integro-differential equation for $g(q, t)$ 
\begin{eqnarray}
\label{9}
\frac{\partial}{\partial t} g(q,t)&=&2\epsilon(1+\cos q)
-2(b-\gamma)g(q,t)+2(a-\gamma)\cos q \;g(q,t)
\nonumber\\&-&\frac{\gamma}{\pi}(1+\cos q)
\int_{-\pi}^{\pi}dk (1-\cos k) g(k,t)
-\frac{2\gamma}{\pi} g(q,t)\int_{-\pi}^{\pi}dk \cos k \; g(k,t) 
\nonumber\\&+&\frac{2\gamma}{\pi}g(q,t)\cos q\int_{-\pi}^{\pi} dk g(k,t)
\end{eqnarray}
with the initial condition $g(q,0)=1$.
This equation with the appropriate boundary conditions 
fully solves the problem of the time evolution of the density and correlation
function. Indeed, introducing the help functions
$F_n(t)=\left(1/ 2\pi\right)\int_{-\pi}^{\pi}dq\; g(q, t)\cos(nq)$ 
(for $n$ integer ), the density can be written as
\begin{eqnarray}
\label{13}
\rho(t)=F_0(t),
\end{eqnarray}
while it is convenient to introduce the function ${\cal G}(r,t)\equiv 1-2\sum_{1\leq n\leq r} F_{2n-r-1}(t)$ to express the (connected) two-point correlation function as
\begin{eqnarray}
\label{14}
{\cal C}_r(t)=\frac{1}{4}\Big[{\cal G}(r-1,t){\cal G}(r+1,t)-{\cal G}^2(r,t)\Big]
\end{eqnarray}
The help functions $F_n(t)$ have the integral representation
\begin{eqnarray}
\label{12}
 F_n(t)&=&e^{-\widetilde K_1(t)}I_n(\widetilde K_0(t))\nonumber\\
&+&\int_{0}^{t}d\tau \Big(\epsilon+\gamma
\{F_1(t-\tau)-F_0(t-\tau)\}\Big)
e^{\widetilde K_1(t-\tau)-\widetilde K_1(t)}
\Bigg\{2I_n\Big(\widetilde K_0(t)-\widetilde K_0(t-\tau))
\nonumber\\&+&I_{n+1}\Big(\widetilde K_0(t)-\widetilde K_0(t-\tau) \Big) 
+ I_{n-1}\Big(\widetilde K_0(t)-\widetilde K_0(t-\tau)\Big) \Bigg\},
\end{eqnarray}
where $I_{n}(z)$ denote the modified Bessel functions of imaginary argument,
and
$\widetilde K_0(t)\equiv 2(a-\gamma)t+4\gamma\int_{0}^{t} ds F_0(s), 
\; \widetilde K_1(t)\equiv 2(b-\gamma)t+4\gamma\int_{0}^{t} ds F_1(s)$.
It is worth noting that $F_0(t)$ and $F_1(t)$ are obtained
self-consistently while the higher order functions
$F_n(t)$ ($n\geq 2$) follow from simple
integration.
Notice that at $\gamma =0$, we recover the results obtained in
\cite{DLSchuetz99,Droz,Grynbergetal,BaresMobilia99,Lushnikov}.
We have solved the integral equations (10) and obtained the $F_n(t)$ 
numerically as well as
analytically in the asymptotic regime. 
In the long time regime, the asymptotes follow as\\
i) In the absence of source ($ht,h't,\epsilon't\gg1\;, \epsilon=0$)
\begin{eqnarray}
\label{15}
\rho(t)&\approx& \frac{1}
{\sqrt{4\pi(h+h')t}}-\frac{\gamma}{\pi\epsilon'(h+h')t}\\
{\cal C}_r(t)&\approx& \frac{-1}{4\pi(h+h')t}
+\Big\{\pi r +\frac{8\gamma}
{\epsilon'}\Big\}\frac{1}{(4\pi(h+h')t)^{3/2}}
\end{eqnarray}
The result for the density agrees with that of reference \cite{Krebsetal}. The numerical computation of $\rho(t)$ is shown in Fig.\ref{fig1}.\\
ii) In the presence of source ($ht,h't,|a|t,bt \gg 1$), 
two regimes have to be distinguished, namely\\

\begin{figure}[b]
 \setlength{\unitlength}{1mm}
  \begin{picture}(80,80)
  \epsfxsize=7cm
 \epsfysize=7cm
 \put(35,0){\epsfbox{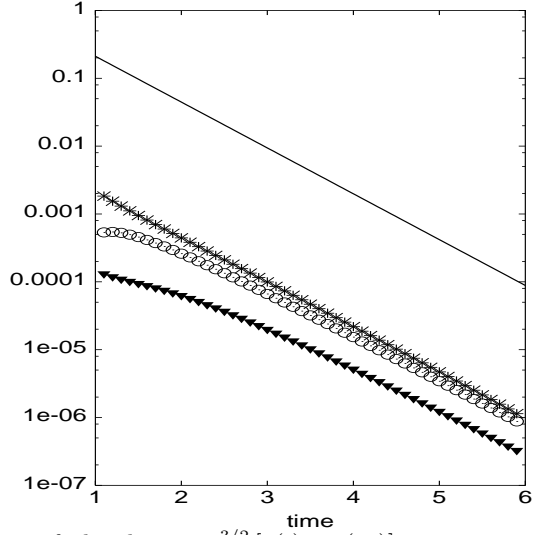}}
  \end{picture}
 \caption{The logarithm of the 
 density $t^{3/2}\left[\rho(t)-\rho(\infty)\right]$ (stars, $\rho(\infty)=1/3$) 
 and of the two-point correlation
 function 
 $-t^{3/2}{\cal C}_r(t),$\, r=1 (circles),\, r=2 (triangles) for $\epsilon=0.5\;, 
 \epsilon'=2$, 
 and $\gamma=1$ versus time. We have included
 the shifted asymptote ${\rm const}-14t/9$ as a guide to the
 eye (thin line). 
 \label{fig2}}
\end{figure}

\noindent
\hskip 0.8cm 1) If ${\widetilde K}_0(t)>0$, then,
\begin{eqnarray}
\label{16}
\rho(t)-\rho(\infty)\sim-{\cal C}_r(t)\sim 
\frac{\exp(-4(\epsilon-\gamma\rho(\infty)^2)t)}{t^{3/2}}
\end{eqnarray}
A numerical illustration of this case is shown in Fig.\ref{fig2}\\

\noindent
\hskip 0.8cm 2) If  ${\widetilde K}_0(t)<0$, then,
\begin{eqnarray}
\label{17}
\rho(t)-\rho(\infty)\sim {\cal C}_r(t)\sim 
\frac{\exp\Big(-2[(b-\gamma)+ 2\gamma F_{1}(\infty)
-|a-\gamma+2\gamma\rho(\infty)|]t\Big)}{t^{1/2}}
\end{eqnarray}
where $\rho(\infty)=1/(1+\sqrt{\epsilon'/\epsilon})$ is the stationary density  which coincides with the (free fermion)
$\gamma=0$ case, and $F_1(\infty)=\sqrt{\epsilon \epsilon'}/
(\sqrt{\epsilon}+\sqrt{\epsilon'})^2$.\\
The results (\ref{15})-(\ref{17}) indicate that the $\gamma$ term
is irrelevant in the long time limit in the sense that it does not renormalize
the exponent of the algebraic (in the absence of source) or 
of the subdominant-algebraic 
(in the presence of source) decay. 
Such a behavior had been observed numerically (in the absence of source) 
in reference \cite{Krebsetal}. 
The correlation times are however renormalized.\\

The extension of our method to deal with epidemic models and many species
is indeed possible. 
Local sources or sinks as well as boundary (initial state) 
terms associated with local operators 
in the hamiltonian formulation can also be dealt with whenever the
Jordan-Wigner phase can be made to cancel.
A physical illustration of this is realized in a model of photoluminescent
diodes with injection currents at the boundaries. 
These and other physically interesting applications of
the method proposed in this paper will be discussed 
elsewhere \cite{BMStochastic99}.
 
% 
%%%%%%%%%%%%%%%%%%%%%%%%%%%%%%%%%%%%%%%%%%%%%%%%%%%%%%%%%%%%%%%%%%%%
%
\section*{ACKNOWLEDGMENTS}
We thank J. Cardy, B. Dou\c{c}ot, M. Droz, Ch. Gruber, Y. Hatsugai, 
Ph. Nozi\`eres, T.M. Rice, G.M. Sch\"utz,  A. Tsvelik and X.-G. Wen 
for discussions. We are especially grateful 
to K. Grzegorczyk, S.Gyger and L.Klinger for computational assistance.
The support of the Swiss National Fonds is gratefully acknowledged.


\end{document}